\documentclass[pra,showpacs,showkeys,amsfonts,amsmath, 11pt]{revtex4}
\usepackage{bm}
\usepackage{graphicx}
\RequirePackage{mathptm}
\newtheorem{thm}{Theorem}[section]

\newtheorem{defn}{Definition}[section]
\newtheorem{rem}{Remark}[section]
\numberwithin{equation}{section}

\newcommand{\ip}[2]{\langle \,{#1},\,{#2}\,\rangle}
\newcommand{\W}[4]{\begin{cases}
#1 ,&#2\\[2mm]
#3 ,&#4
\end{cases}}

\newcommand{\om}{\omega}

\newcommand{\vf}{\varphi}

\newcommand{\I}{\openone}
\newcommand{\conj}[1]{\overline{#1}}

\newcommand{\cH}{{\mathcal H}}

\newcommand{\C}{\mathbb C}
\newcommand{\R}{\mathbb R}

\newcommand{\N}{\mathbb N}

\newcommand{\Hb}{\mathbb H}

\newcommand{\Gb}{\mathcal G}
\newcommand{\Wu}{\mathfrak W}
\newcommand{\Mu}{\mathfrak M}
\newcommand{\fA}{\mathfrak A}

\newcommand{\cA}{{\mathcal A}}

\newcommand{\cF}{{\mathcal F}}

\newcommand{\mr}[1]{\mathrm{#1}}

\newcommand{\bh}{\mathbf{h}}

\begin{document}
\title{Variable Planck's constant and scaling properties of states on Weyl algebra}
\author{Piotr {\L}ugiewicz \footnote{piotr.lugiewicz@ift.uni.wroc.pl}, Lech Jak{\'o}bczyk \footnote{ lech.jakobczyk@ift.uni.wroc.pl} and Andrzej Frydryszak \footnote{ andrzej.frydryszak@ift.uni.wroc.pl}}
 \affiliation{Institute of Theoretical Physics\\ University of
Wroc{\l}aw\\
Plac Maxa Borna 9, 50-204 Wroc{\l}aw, Poland}
\begin{abstract}
We consider the possible quantum effect for infinite systems produced  by variations of the Planck's constant.
Using the algebraic formulation of quantum theory we study behaviour of states $\omega$ defined as positive, normalized
functionals on the canonical commutation relations algebra (CCR-algebra) under the changes of the defining relations of the
CCR. These defining relations of the multiplication in the CCR-algebra depend explicitly on the value of the Planck's constant.
We analyse to what extend changes of the $\hbar$ preserve the original state space (this gives restrictions on the admissible
changes of the Plank's constant) and what  properties have original quantum states $\omega$  as states on the new algebra.
We answer such questions for the quasi-free states.  We show that any universally invariant state can be interpreted as a
convex combination of Fock states with different values of Planck's constant.   The second important class of states we study
are the KMS-states, here the rescaling alters in a nontrivial way the relevant dynamics.
We also show that it is possible to go beyond the limits restricting the changes of the $\hbar$, but then one has to restrict the CCR-algebra to a subalgebra.
\end{abstract}
 \maketitle
\section{Introduction}
One can observe recently mounting interest in the fundamental question of the possible variability of the Planck's
constant $\hbar$ \cite{Dirac, Uzan1, Uzan2, Webb, Webb-sp, Mangano, Hutchin}. As it sounds paradoxical, questioning of the absolutness of the $\hbar$ as a fundamental constant entity and considering it as
"material" coefficient \cite{Hutchin} characterizing locally the quantumness of the fabric of space-time, is of principal interest. Dirac's suggestion that the fundamental constants can depend on the epoch of the Universe \cite{Dirac} inspired many authors. As already existing works indicate, it touches not only the questions concerning models of the Universe including with dark energy and matter, its beginings, but also experiments performed locally
at human scales. The topic is around for a long time, formerly scattered in isolated research, it has grown into a more definite flow of works as theoretical as experimental. It is worth noting that some experiments verifying variability of the Planck's constant are already run for decades \cite{Hutchin}. As spatial as temporal \cite{Mangano} variations of $\hbar$ are taken into consideration. Possible influences of such variations on macroscopic world are being considered \cite{Yang}. On the other hand, a formal variability of the $\hbar$ has always been accepted in the delicate context of the so called classical limit for given quantum system, or commutative limit of canonical commutation relations within the canonical quantization scheme. An example of strict formalism, where the Planck's constant is related to a running parameter controlling the non-commutativity, is the formalism of the (strict) deformation quantization (see e.g. \cite{Rieffel, Bordemann, Landsman-jmp, Landsman-book}), where classical limit is literarily performed as the vanishing $\hbar$ limit.
\par
Recently, M.A. de Gosson \cite{Gosson} considered quantum mechanical
consequences of possible changes of Planck's constant. Using the
Wigner formalism this author shows that the purity of the state is
extremely sensitive to such changes. In our article we study the
similar problems, but in the case of quantum systems with infinite
number of degrees of freedom, such as quantum fields or quantum
statistical systems. In the algebraic setting, possible changes of
Planck's constant which influence  commutation relations of basic
objects, modify the structure of the algebra of canonical
commutation relations (CCR). In particular, in the Weyl form of CCR,
the product of elements of the algebra is defined in terms of
symplectic form $\sigma$ on the one - particle space $\Hb$, which
manifestly depends on the Planck's constant. In the algebraic
formulation of quantum theory, the states are identified with
positive and normalized functionals $\om$ on the corresponding.
algebra of CCR, so the problem of sensitivity of quantum states to
the changes of Planck's constant can be formulated as follows.
Suppose that $\om$ is the state on the CCR algebra given by the Planck's constant
 fixed at its "standard" value. Consider now the
same functional $\om$ but as a state on the algebra with changed Planck's
constant i.e. with changed multiplication law. In this context,
the natural questions arise: \\
(i) under which condition this is possible?\\
(ii) what are the properties of $\om$ considered as a state on the
new algebra?\\
In the  present paper we consider these  questions in the case of
quasi - free states. To simplify the analysis we fix standard Planck's constant $\hbar=1$
and possible changes of Planck's constant are measured by dimensionless scaling parameter $h$. The starting point is the CCR algebra
defined by symplectic form $\sigma$ with $h=1$, denoted by $\Wu(\Hb,\sigma)$. The algebras corresponding to changed Planck's constant
 are defined in terms of rescaled symplectic forms $\sigma_{h}$ and are denoted by $\Wu(\Hb,\sigma_{h})$. We show that given state $\om$ on the algebra $\Wu(\Hb,\sigma)$ can be also considered as a state on $\Wu(\Hb,\sigma_{h})$ if a properly defined rescaled functional $\om_{h}$ is a state on the algebra $\Wu(\Hb,\sigma)$. It turns out that this condition  gives a bound on the admissible values of the scaling parameter $h$.
 On the other hand, the rescaled state $\om_{h}$ on $\Wu(\Hb,\sigma)$ can be treated as isomorphic image of the state $\om$ on the algebra $\Wu(\Hb,\sigma_{h})$, under the natural $\ast$ - isomorphism between the algebras $\Wu(\Hb,\sigma)$ and $\Wu(\Hb,\sigma_{h})$.
 \par
 The study of the properties of these rescaled states is the main purpose of the present work. The first result shows that for any quasi-free state $\om$ defined by some operator $A\geq \I$, the rescaled functional $\om_{h}$ is a quasi - free state on $\Wu(\Hb,\sigma)$, provided the scaling parameter $h$ is not larger that the bottom of the spectrum of $A$. In the case of Fock state $\om_{0}$ where $A=\I$, $0<h<1$ and all rescaled states $\om_{0,h}$ are non - Fock states with non - zero expectation value of the one - particle number operator. On the other hand, the total number operator does not exists, since the states $\om_{0,h}$ are not quasi - equivalent to the Fock state $\om_{0}$. The states $\om_{0,h}$ have another interesting property. These states are extreme points of the convex set of all universally invariant states of the algebra $\Wu(\Hb,\sigma)$ (see Sect. IIIC). So any universally invariant state can be interpreted as a convex combination of Fock states with different values of Planck's constant. Next we study the properties of equilibrium states. We adopt the general definition in terms of Kubo - Martin - Schwinger condition. Again the changes of Planck's constant alter the properties of states, but as we show the state $\om$ being KMS - state at the inverse temperature $\beta$, after rescaling remains $\beta$ - KMS state but with respect to the dynamics altered in a quite complicated way. The new hamiltonian is affiliated to the algebra generated by the hamiltonian defining the starting dynamics of the system. When the scaling parameter $h$ is beyond the admissible set of values, in particular $h$ is larger then the bottom of the spectrum of $A$, the quasi - free states $\om_{h}$ have remarkable properties. It turns out that $\om_{h}$ are positive - definite only after restriction to some subalgebra of the Weyl algebra $\Wu(\Hb,\sigma)$. On the other hand, such restricted states can be extended to some  non - regular states $\widetilde{\om}_{h}$ on the whole Weyl algebra. It is worth to stress that for some physically interesting systems (especially in connection with quantum gauge theories) the lack of regularity of states is not a mathematical pathology, but has interesting theoretical consequences (for detailed discussion see \cite{Strocchi}).
  When the scaling parameter achieves its maximal value $h_{\ast}=||A||_{\mr{op}}$, the corresponding non - regular states are all equal to the unique trace - state $\om_{\infty}$ i.e. the state satisfying $\om_{\infty}(AB)=\om_{\infty}(BA)$ for all $A,B \in \Wu(\Hb,\sigma)$. Such a state, formally can be interpreted as the equilibrium state at infinite ($\beta =0$) temperature.

\section{Algebra of commutation relation and states}
\subsection{Weyl algebra of canonical commutation relation}
A Weyl algebra of canonical commutation relations usually is constructed in terms of exponentials of canonical field operators
$\Phi(f)$, with $f$ belonging to  real vector space $\Hb$ endowed with symplectic form $\sigma(f,g)$, which encodes CCR relations (we put $\hbar=1$)
\begin{equation}
[\Phi(f),\Phi(g)]=i\, \sigma(f,g)
\end{equation}
If the operators $\Phi(f)$ are self - adjoint, we define
\begin{equation}
W_{f}=e^{i\Phi(f)}
\end{equation}
and obtain the relations
\begin{equation}
W_{f}W_{g}=e^{-\frac{i}{2}\sigma(f,g)}W_{f+g}
\end{equation}
Generally, we define an abstract CCR algebra in the Weyl form \cite{MSTV} as follows.
Let $\Hb$ be a real linear space and $\sigma$ a symplectic form on
$\Hb$. We consider also symplectic forms $\sigma_{h},\; h\in \R,\;
h>0$ as
$$
\sigma_{h}(f,g)=h\,\sigma (f,g),\quad f,g \in \Hb
$$
Let $W_{f}$ be the function on $\Hb$ given by
$$
W_{f}(g)=\W{1}{f=g}{0}{f\neq g}
$$
Let $\Wu_{0}(\Hb,\sigma_{h})$ be the $\ast$ - algebra generated by
$\{W_{f}\,:\, f\in \Hb\}$. The elements $W_{f}$ satisfy
\begin{equation}
W_{f}\,W_{g}=e^{-\frac{i}{2}\sigma_{h}(f,g)}\,W_{f+g}
\end{equation}
and
\begin{equation}
W_{f}^{\ast}=W_{-f}
\end{equation}
On the algebra $\Wu_{0}(\Hb,\sigma_{h})$ we define the norm
\begin{equation}
||\sum\limits_{j=1}^{n}a_{j}\,\delta_{f_{j}}||_{1}=\sum\limits_{j=1}^{n}|a_{j}|
\end{equation}
and let $\conj{\Wu_{0}(\Hb,\sigma_{h})}^{||\cdot||_{1}}$ be the completion of
$\Wu_{0}(\Hb,\sigma_{h})$ with respect to this norm. So defined
algebra is $\ast$ - Banach algebra. To obtain C$^{\ast}$ - algebra
of CCR one introduces minimal regular norm
\begin{equation}
||A||=\sup\limits_{\pi}||\pi(A)||,\quad A\in \conj{\Wu_{0}(H,\sigma_{h})}^{||\cdot||_{1}}
\end{equation}
where $\pi$ are $\ast$ - representations of $\conj{\Wu_{0}(\Hb,\sigma_{h})}^{||\cdot||_{1}}$.
The completion of $\conj{\Wu_{0}(\Hb,\sigma_{h})}^{||\cdot||_{1}}$ with respect to this norm is
C$^{\ast}$ - algebra of CCR denoted by $\Wu(\Hb,\sigma_{h})$.
\begin{thm}
For any $h>0$ there exists a unique $\ast$ - isomorphisms
$\gamma_{h}$ from $\Wu(\Hb,\sigma_{h})$ onto $\Wu (\Hb,\sigma)$.
\end{thm}
\textit{Proof:} We follow here the arguments from \cite{HR}. Let us
denote the Weyl elements generating the algebra
$\Wu_{0}(\Hb,\sigma_{h})$ by $W_{f}^{h}$. Define
\begin{equation}
\gamma_{h}(W_{f}^{h})=W^{h=1}_{\sqrt{h}f},\quad f\in \Hb\label{iso}
\end{equation}
Linear extension of Eq. (\ref{iso}) gives a bijection $\gamma_{h}$
from $\Wu_{0}(\Hb,\sigma_{h})$ onto $\Wu_{0}(\Hb,\sigma)$. Since
$$
W_{f}^{h\;\ast}=W^{h}_{-f}
$$
we have
$$
\gamma_{h}(W_{f}^{h})^{\ast}=W_{\sqrt{h}f}^{\ast}=\gamma_{h}(W^{h\;\ast}_{f})
$$
which by linear extension leads to the relation
$$
\gamma_{h}(A)^{\ast}=\gamma_{h}(A^{\ast}),\quad A\in
\Wu_{0}(\Hb,\sigma_{h})
$$
Since
$$
\gamma_{h}(W_{f}^{h}W_{g}^{h})=e^{-\frac{i}{2}h\, \sigma
(f,g)}\,\gamma_{h}(W^{h}_{f+g})=e^{-\frac{i}{2}\sigma
(\sqrt{h}f,\sqrt{h}g)}W_{\sqrt{h}f+\sqrt{h}g}=\gamma_{h}(W^{h}_{f})\gamma_{h}(W^{h}_{g})
$$
it follows that  for all $A,B\in \Wu_{0}(\Hb,\sigma_{h})$
$$
\gamma_{h}(AB)=\gamma_{h}(A)\gamma_{h}(B)
$$
and $\gamma_{h}$ is $\ast$ - isomorphisms from
$\Wu_{0}(\Hb,\sigma_{h})$ onto $\Wu_{0}(\Hb,\sigma)$, which can be
extended by continuity.
\subsection{States on $\Wu(\Hb, \sigma_{h})$}
A state on $\Wu(\Hb,\sigma_{h})$ is normalized and positive linear functional.
\begin{defn}
The function $\vf\,:\, \Hb\to \C$ is $\sigma_{h}$ - positive if for
all $n\in \N$ and $a_{k}\in \C,\; f_{k}\in \Hb;\: k=1,\ldots,n$
\begin{equation}
\sum\limits_{j,k=1}^{n}a_{j}\conj{a}_{k}e^{-\frac{i}{2}\sigma_{h}(f_{j},f_{k})}\,\vf(f_{j}-f_{k})\geq
0\label{sigmahpositive}
\end{equation}
\end{defn}
We say that $\vf$ is $\sigma$ - positive if it is $\sigma_{h}$ -
positive for $h=1$. The set of all $\sigma_{h}$ - positive functions
on $\Hb$ will be denoted by $\cF_{h}$. We write $\cF_{1}=\cF$.
Define
$$
\vf_{h}(f)=\vf(\frac{1}{\sqrt{h}}f)
$$
\begin{thm}
$$
\vf\in \cF_{h}\quad \Leftrightarrow\quad \vf_{h}\in \cF
$$
\end{thm}
\textit{Proof:} If $\vf\in \cF_{h}$, then condition
(\ref{sigmahpositive}) is satisfied. Now the lefthand side of
(\ref{sigmahpositive}) can be written in the following way
$$
\sum\limits_{j,k=1}^{n}a_{j}\conj{a}_{k}e^{-\frac{i}{2}\sigma(\sqrt{h}f_{j},\sqrt{h}f_{k})}\,\vf(f_{j}-f_{k})=
\sum\limits_{j,k=1}^{n}a_{j}\conj{a}_{k}e^{-\frac{i}{2}\sigma(\tilde{f}_{j},\tilde{f}_{k})}\,
\vf_{h}(\tilde{f}_{j}-\tilde{f}_{k})
$$
where $\tilde{f}_{j}=\sqrt{h}f_{j}$. It means that the function
$\vf_{h}\in \cF$. Similarly we show that if $\vf_{h}$ is $\sigma$ -
positive, then $\vf$ is $\sigma_{h}$ - positive.  The following
result is standard:
\begin{thm}
The functional $\om$ is a state on $\Wu(\Hb,\sigma_{h})$ if and only if the
generating function of $\om$ i.e the function
$\vf_{\om}(f)=\om(W_{f}),\; f\in \Hb$, belongs to $\cF_{h}$.
\end{thm}
Fix some value of the parameter $h$ and consider the state $\om$ on
CCR algebra $\Wu(\Hb,\sigma_{h})$. $\om$ is a linear functional on
$\ast$ - algebra generated by elements $W_{f},\; f\in \Hb$. By
definition this functional is positive - definite, when the product
of elements of the algebra is defined in terms of symplectic form
$\sigma_{h}$, so $\om$ can be extended to positive - definite
functional on C$^{\ast}$ - algebra $\Wu(\Hb,\sigma_{h})$. The
corresponding generating function $\vf_{\om}$ is $\sigma_{h}$ -
positive. Notice that if $\tilde{h}\neq h$, the same functional can
be considered on the algebra $\Wu_{0}(\Hb,\sigma_{\tilde{h}})$ where
the multiplication is differently defined. But it can happen that
$\om$ is still positive - definite, so in this case the state $\om$
can be extended to C$^{\ast}$ - algebra
$\Wu(\Hb,\sigma_{\tilde{h}})$. In particular we have
\begin{thm}
Let $\om$ be a state on CCR algebra $\Wu(\Hb,\sigma)$. The
functional $\om$ can also be considered as a state on the algebra
$\Wu(\Hb,\sigma_{h})$ if the rescaled functional $\om_{h}$ given
by
\begin{equation}
\om_{h}=\om \circ \tau_{h},\quad
\tau_{h}(W_{f})=W_{(1/\sqrt{h})f}\label{omegah}
\end{equation}
is a state on the algebra $\Wu (\Hb,\sigma)$.
\end{thm}
\begin{rem}
Notice that $\tau_{h}=\gamma_{h}^{-1}$ and rescaled state $\om_{h}$ is the isomorphic image of the functional $\om$ considered as a state on the algebra $\Wu(\Hb,\sigma_{h})$.
\end{rem}
\section{Quasi - free states}
\subsection{ General definition}
In the following we consider the case when $\Hb$ is a complex, infinite - dimensional
Hilbert space with a scalar product $\ip{\cdot}{\cdot}$ and $\sigma
(f,g)= \mathrm{Im}\, \ip{f}{g}$. Let $S(f,g)$ be a positive sesqulinear form on $\Hb$. Assume also that $S(f,g)$ is bounded.
Then there exists a bounded linear operator $A$, such that
\begin{equation}
S(f,g)=\ip{f}{Ag}\label{sA}
\end{equation}
The function $\vf_{S}$ defined by
\begin{equation}
\vf_{S}(f)=e^{-\frac{1}{4}\, S(f,f)}\label{sform}
\end{equation}
is a generating function of some state $\om$ on the algebra $\Wu(\Hb,\sigma)$, if it
is $\sigma$ - positive.  One can show (see e.g. \cite{Petz}) that  $\sigma$ - positivity of (\ref{sform}) is equivalent to the following condition
\begin{equation}
|\sigma (f,g)|^{2}\leq S(f,f)S(g,g)\label{sigmascond}
\end{equation}
 In the case when the form $S(f,g)$ is defined by bounded operator $A$, the condition (\ref{sigmascond})
is satisfied if and only if
\begin{equation}
A\geq \I\label{A}
\end{equation}
So if (\ref{A}) is satisfied, the functional
\begin{equation}
\om(W_{f})=\vf_{S}(f)\label{QFS}
\end{equation}
extended by linearity and continuity to the whole $\Wu(\Hb)$ is so called \textit{gauge - invariant quasi - free state} on CCR algebra $\Wu(\Hb,\sigma)$.
\par
\par
The explicit construction of GNS representation $(\pi_{\om},\cH_{\om},\Omega_{\om})$ of the Weyl algebra $\Wu (\Hb,\sigma)$ defined by quasi - free state (\ref{QFS}), can be given as follows \cite{MRT}: Let $\cH=\Gamma (\Hb)$ be the bosonic Fock space over
$\Hb$. Define
\begin{equation}
T_{1}=\frac{1}{\sqrt{2}}(A+\I)^{1/2},\quad
T_{2}=\frac{1}{\sqrt{2}}(A-\I)^{1/2}
\end{equation}
GNS representation induced by
$\om$ is unitary equivalent to the representation defined on
$\cH_{\om}=\cH\otimes \cH$ and
\begin{equation}
\pi_{\om}(W_{f})=\pi_{0}(W_{T_{1}f})\otimes \pi_{0}(W_{T_{2}f})
\end{equation}
where $\pi_{0}$ is the Fock representation on $\cH$. One can check
that $\Omega_{\om}=\Omega_{0}\otimes \Omega_{0}$ (where $\Omega_{0}$ is the
vacuum vector in $\cH$) is a cyclic vector for $\pi_{\om}$. The
representation $\pi_{\om}$ is reducible, in fact $\tilde{\pi}_{\om}$
given by
\begin{equation}
\tilde{\pi}_{\om}(W_{f})=\pi_{0}(W_{T_{2}f})\otimes
\pi_{0}(W_{T_{1}f})
\end{equation}
commutes with $\pi_{\om}$ and has $\Omega_{\om}$
as a cyclic vector, too. So $\Omega_{\om}$ is cyclic and
separating for the von Neumann algebra $\Mu_{\om}$ generated by
operators $\pi_{\om}(W_{f})$.
\par
The quasi - free state $\om$ is regular, or even analytic i.e. the mapping $t\to \om(W_{tf})$ for all  $f\in \Hb$
is analytic in an open neighborhood of the origin. For each $f\in\Hb$, denote by $\Phi_{\om}(f)$ the infinitesimal generator of the unitary group
$t\to \pi_{\om}(W_{tf})$. Define also the annihilation and creation operators, by
\begin{equation}
D(a_{\om}(f))=D(a_{\om}^{\ast}(f))=D(\Phi_{\om}(f))\cap D(\Phi_{\om}(if))
\end{equation}
and
\begin{equation}
a_{\om}(f)=\frac{1}{\sqrt{2}}\left(\Phi_{\om}(f)+i\,\Phi_{\om}(if)\right),\quad a_{\om}^{\ast}=\frac{1}{\sqrt{2}}\left(\Phi_{\om}(f)-i\,\Phi_{\om}(if)\right)
\end{equation}
The operators $a_{\om}(f),\; a_{\om}^{\ast}(f)$ are densely defined, closed and $a_{\om}(f)^{\ast}=a_{\om}^{\ast}(f)$. Let us introduce the self - adjoint operator
\begin{equation}
N_{f}= a_{\om}^{\ast}(f)a_{\om}(f)
\end{equation}
We take $N_{f}$ as a number operator for the one particle state $f\in \Hb$. One can check that
\begin{equation}
\om(N_{f})=\om(a_{\om}^{\ast}(f)a_{\om}(f))=\frac{1}{2}\,\ip{f}{(A-\I)f}
\end{equation}
The definition of total number operator is much more involved (see e.g. \cite{BR}). One can proceed as follows. The finite - dimensional
subspaces $F\subseteq \Hb$ form a directed set when ordered by inclusion. If $\{f_{j}\}$ is an orthonormal basis for $F$, define
$$
D(n_{\om, F})=\bigcap\limits_{f_{j}\in F}D(a_{\om}(f_{j}))
$$
and
$$
n_{\om,F}(\Psi)=\sum\limits_{f_{j}\in F}||a_{\om}(f_{j})||^{2}
$$
The set $\{n_{\om,F}\}$ forms monotonically increasing net of positive, closed quadratic forms. Their limit will be also positive, closed quadratic form on $\cH_{\om}$. If this form is densely defined, it determines a unique self - adjoint total number operator $N_{\om}$ on the Hilbert space $\cH_{\om}$ and one can show \cite{CMR} that $N_{\om}$ exists when the state is quasi - equivalent to the Fock state $\om_{0}$ i.e. quasi - free state with $A=\I$. More precisely, it means that the GNS representation $\pi_{\om}$ is unitary equivalent to direct sum of copies of the Fock representation.
\subsection{Rescaled quasi - free state $\om_{h}$}
Now we would like to study the properties of $\omega$ considered as
functional defined on  CCR algebra $\Wu(\Hb,\sigma_{h})$. To this
end we consider the properties of rescaled state $\om_{h}$. In
this case the state $\om_{h}$ is defined by the function
$$
\vf_{S}^{h}(f)=(\om\circ
\tau_{h})(W_{f})=e^{-\frac{1}{4}\ip{f}{A_{h}f}},\quad
A_{h}=\frac{1}{h}A
$$
which belongs to $\cF$ if
$$
\frac{1}{h}\,A\geq \I
$$
Obviously this condition is satisfied if $h\in (0,\; 1)$. More generally,
$$
h\leq h_{\mr{max}}=\inf\, \mr{spec}(A)
$$
\begin{thm}
The quasi - free state $\omega$ defined by the operator $A\geq \I$
can be considered as a quasi - free state on  all algebras
$\Wu(\Hb,\sigma_{h})$, for $h\in (0,\,h_{\mr{max}}]$, where
$h_{\mr{max}}=\inf\, \mr{spec}(A)$. In particular, for all such
values of the parameter $h$, the rescaled state $\om_{h}$ is a
quasi - free state on the algebra $\Wu (\Hb, \sigma)$.
\end{thm}
\subsection{Rescaled Fock state}
The simplest quasi - free state on the algebra $\Wu(\Hb,\sigma)$ is
the Fock state $\om_{0}$ given by $A=\I$. This state is pure and it
means that GNS representation corresponding to $\om_{0}$ is
irreducible (see e.e. \cite{Petz}). Now we consider the rescaled
functional $\om_{0,h}$. This functional will be positive -
definite for all  $h\in (0,1)$  and can be written as
\begin{equation}
\omega_{0,h}(W_{f})=e^{-\frac{1}{4}\,\ip{f}{A_{h} f}},\quad
A_{h}=\frac{1}{h}\I
\end{equation}
so
\begin{equation}
\om_{0,h}(W_{f})=e^{-\frac{1}{4}||f||^{2}(1/h)}
\end{equation}
and we have
\begin{thm}
The rescaled Fock state $\om_{0,h}$ on the CCR algebra
$\Wu(\Hb,\sigma)$ is the quasi - free (non - Fock) state for all
$0<h<1$.
\end{thm}
Notice that in the state $\om_{0,h}$, the number operator $N_{f}$ has non - zero expectation value
$$
\om_{0,h}(N_{f})=\frac{1-h}{2h}\ip{f}{f}=\frac{1-h}{2h}
$$
but the total number operator does not exists, since the following theorem is true:
\begin{thm}
The state $\om_{0,h}$ is not quasi - equivalent to the Fock state $\om_{0}$.
\end{thm}
\textit{Proof:} The proof is based on the standard result: quasi - free state defined by operator $A\geq \I$ is quasi - equivalent to the Fock state
if and only if $A-\I$ has finite trace (see e.g. \cite{C}). In the case of the state $\om_{0,h}$ we have
$$
A_{h}-\I=\frac{1-h}{h}\,\I
$$
which in the case of infinite dimensional space $\Hb$ is not trace  class.
\par
It is worth to notice that re scaled Fock states $\om_{0,h}$ are in fact the extreme points of the convex set of all universally invariant states \cite{Segal}. To define this notion, let
$$
\alpha_{U}(W_{f})=W_{Uf}
$$
for any unitary operator on the Hilbert space $\Hb$.
The state $\om$ on the algebra $\Wu(\Hb,\sigma)$ is universally invariant if for all unitary $U$
$$
\om \circ \alpha_{U}=\om
$$
It can be shown that any regular state $\om$ on the Weyl algebra $\Wu(\Hb,\sigma)$ which is universally invariant has the form
\begin{equation}
\om=\int\limits_{0}^{1}\om_{c}\,dm(c)
\end{equation}
where $m$ is a probability measure on $[0,1)$ and the state $\om_{c}$ is given by
\begin{equation}
\om_{c}(W_{f})=e^{-\frac{1}{4}||f||^{2}((1+c)/(1-c))}
\end{equation}
so
\begin{equation}
\om_{0,h}=\om_{c}\quad\text{for}\quad c=\frac{1-h}{1+h}
\end{equation}
\section{Equilibrium states}
\subsection{KMS state on the algebra $\Wu(\Hb,\sigma)$}
We start with the general definition of the equilibrium state by the  Kubo - Martin - Schwinger (KMS) condition \cite{BR}. Let
$\fA$ be the C$^{\ast}$ - algebra of observables and $\{\alpha_{t}\}_{t\in\R}$ be the one - parameter group of $\ast$ - atomorphisms
of $\fA$ describing time evolution. The state $\om$  is $(\alpha_{t},\,\beta)$ - KMS state  at the inverse
temperature $\beta$ if for any pair $A,\, B\in \fA$, there exists a complex function $f_{A,B}(z)$ which is analytic on the strip
$$
{\mathcal T}_{\beta}=\{z\,:\, 0<\mr{Im}z<\beta\}
$$
and bounded and continuous on $\conj{{\mathcal T}_{\beta}}$, such that
$$
f_{A,B}(t)=\om  (A,\alpha_{t}(B)),\quad f_{A,B}(t+i\beta)=\om (\alpha_{t}(B)A)
$$
for all $t\in \R$.
\par
Now we apply this definition to he case of quasi - free state on the
CCR algebra $\Wu(\Hb,\, \sigma)$ defined by the operator $A\geq \I$.
Let $T_{t}$ be a group of unitary operators on $\Hb$ with self -
adjoint generator $\bh$.  Assume that the form $S(f,g)=\ip{f}{Ag}$
is $T_{t}$ - invariant and define
$$
\alpha_{t}(W_{f})=W_{T_{t}f}
$$
 For quasi - free states, two - point correlation functions
$$
\Gb(f,g\,;\,t)=\omega(W_{f}W_{T_{t}g})
$$
have the form
$$
\Gb(f,g\,;\,t)=e^{\frac{1}{4}S(f,f)-\frac{1}{4}S(g,g)}
\,e^{-\frac{1}{2}F(f,g\,;\,t)}
$$
where
$$
F(f,g\,;\,t)=\mr{Re}\,\ip{f}{T_{t}Ag}+i\; \mr{Im}\ip{f}{T_{t}g}
$$
Now  $\om$ is $(\alpha_{t},\,\beta)$ -KMS with state
if  for every pair $f,g\in \Hb$ there exists
a function $\Phi(f,g\,;\, z)$, analytic on the strip $\mathcal{T}_{\beta}$ and continuous on the boundary, such that
$$
\Phi(f,g\,;\, t+i\,0)=F(f,g\,;\,t)\quad\text{and}\quad \Phi(f,g\,;\,
t+i\beta)=F(g,f\,;\,-t)
$$
It can be proved \cite{GJO} that above condition implies that $\bh \geq \epsilon \I$ for $\epsilon >0$ and
\begin{equation}
A=\frac{\I+e^{-\beta \bh}}{\I-e^{-\beta\bh}}
\end{equation}
Observe that
$$
\mr{spec}(A)\subset [1,\, ||A||_{\mr{op}}]
$$
where
$$
||A||_{\mr{op}}=\frac{e^{\beta \epsilon}+1}{e^{\beta\epsilon}-1}
$$
Notice also that
\begin{equation}
F(f,g\,;\,
t)=\frac{1}{2}\ip{f}{e^{it\bh}(A+\I)g}+\frac{1}{2}\ip{g}{e^{-it\bh}(A-\I)f}\label{F}
\end{equation}
Let us define positive operator $\Delta=e^{\bh}$ on the canonical
domain $D$. Then
\begin{equation}
\Delta= \left(\frac{A+\I}{A-\I}\right)^{1/\beta}\label{Delta}
\end{equation}
and
\begin{equation}
T_{t}=\Delta^{it}=\left(\frac{A+\I}{A-\I}\right)^{it/\beta}
\end{equation}
Observe that
$$
A-\I=(A+\I)\Delta^{-\beta}
$$
so the operators
$$
(A-\I)\Delta^{-iz},\quad z\in \conj{\mathcal{T}_{\beta}}
$$
are bounded. Obviously  operators $(A+\I)\Delta^{iz},
\; z\in \conj{\mathcal{T}_{\beta}}$ are bounded, so for all pairs of vectors $f,\, g\in \Hb$ we obtain the explicit formula for the analytic function
$\Phi(f,g\,;\, z)$:
\begin{equation}
\Phi(f,g\,;\,
z)=\frac{1}{2}\ip{f}{(A+\I)\Delta^{iz}g}+\frac{1}{2}\ip{g}{(A-\I)\Delta^{-iz}f}\label{Phi}
\end{equation}
\subsection{Rescaled KMS state $\om_{h}$}
In this subsection we study the properties of rescaled  KMS state
$\omega_{h}$. It is natural to assume that one - particle
hamiltonian $\bh$ is unbounded, so
\begin{equation}
\inf\, \mr{spec}(A)=1
\end{equation}
and  scaling parameter $h \in (0,1)$. Now the state $\omega_{h}$ is
defined by the operator
\begin{equation}
A_{h}=\frac{1}{h}A\label{Ah}
\end{equation}
which still satisfies $A_{h}\geq \I$.
In the analogy  to the case $h=1$, we define
\begin{equation}
\Delta_{h}=\left(\frac{A_{h}+\I}{A_{h}-\I}\right)^{1/\beta}\label{Deltah}
\end{equation}
or more precisely
\begin{equation}
\Delta_{h}=\int\limits_{e^{\epsilon}}^{\infty}j_{h}(\lambda)\,dP(\lambda)
\end{equation}
on the canonical domain  $D^{h}$, where
\begin{equation}
j_{h}(\lambda)=\left(\frac{1-h+(1+h)\lambda^{\beta}}{1+h+(1-h)\lambda^{\beta}}\right)^{1/\beta},\quad
\lambda\in [e^{\epsilon},\infty)\label{jh}
\end{equation}
and $dP(\lambda)$ is the spectral measure of the operator $\Delta$.  The function (\ref{jh}) satisfies
\begin{equation}
j_{h}(\lambda)\geq j_{h}(e^{\epsilon})=e^{\delta}>1\label{ec}
\end{equation}
 We define also the unitary group
\begin{equation}
T_{t}^{h}=\Delta_{h}^{it}=\left(\frac{A_{h}+\I}{A_{h}-\I}\right)^{it/\beta}
\end{equation}
By $\bh_{h}$ we denote the generator of this group. The operator
$\bh_{h}$  can be  recovered from $\Delta_{h}$ by the formula
\begin{equation}
\bh_{h}=\log \,\Delta_{h}
\end{equation}
Notice that the bottom of the spectrum of the operator $\bh_{h}$ is given by the constant $\delta$ in the formula (\ref{ec}).
Define the function
\begin{equation}
F^{h}(f,g\,;\,t)=\frac{1}{2}\ip{(A_{h}+\I)f}{\Delta_{h}^{it}g}+\frac{1}{2}\ip{(A_{h}-\I)g}{\Delta_{h}^{-it}f}
\end{equation}
Using the same arguments as before, we show that the functions
\begin{equation}
\Phi^{h}(f,g\,;\,
z)=\frac{1}{2}\ip{f}{(A_{h}+\I)\Delta_{h}^{iz}g}+\frac{1}{2}\ip{g}{(A_{h}-\I)\Delta_{h}^{-iz}f}\label{Phih}
\end{equation}
are analytic on the strip $\mathcal{T}_{\beta}$ and continuous on $\conj{\mathcal{T}_{\beta}}$.
Moreover
\begin{equation}
\Phi^{h}(f,g\,;\,t+i\,0)=\frac{1}{2}\ip{(A_{h}+\I)f}{\Delta_{h}^{it}g}+\frac{1}{2}\ip{(A_{h}-\I)g}{\Delta_{h}^{-it}f}
=F^{h}(f,g\,;\,t)\label{Fh0}
\end{equation}
and
\begin{equation}
\begin{split}
\Phi^{h}(f,g\,;\,t+i\beta)&=\frac{1}{2}\ip{\Delta^{-it}_{h}\Delta^{-\beta}_{h}f}{(A_{h}+\I)g}
+\frac{1}{2}\ip{(A_{h}-\I)g}{\Delta_{h}^{-it}\Delta_{h}^{\beta}f}\\
&=\frac{1}{2}\ip{f}{(A_{h}-\I)\Delta_{h}^{it}g}+\frac{1}{2}\ip{g}{(A_{h}+\I)\Delta^{-it}f}\\
&=F^{h}(g,f\,;\,-t)
\end{split}\label{Fhbeta}
\end{equation}
Existence of the analytic functions (\ref{Phih}) satisfying boundary
conditions (\ref{Fh0}) and (\ref{Fhbeta}) means the state $\om_{h}$
is $(\alpha_{t}^{h},\,\beta)$ - KMS state on the algebra
$\Wu(\Hb,\sigma)$, where
\begin{equation}
\alpha_{t}^{h}(W_{f})=W_{T_{t}^{h}f}
\end{equation}
So we arrive at the result
\begin{thm}
Let $h\in (0,1)$ and $\om$ is  $(\alpha_{t},\,\beta)$ - KMS state on
the algebra $\Wu(\Hb,\sigma)$. The rescaled state $\om_{h}$ is
$(\alpha_{t}^{h},\, \beta)$ - KMS state on the algebra $\Wu(\Hb,\sigma)$.
 \end{thm}
\section{Beyond the admissible values of the scaling parameter}
As we have shown, the admissible values of the scaling parameter $h$ in the case of quasi - free state belong to the interval $(0, h_{\mr{max}})$, where $h_{\mr{max}}=\inf\, \mr{spec}(A)$. In this Section we study the properties of rescaled quasi - free states for the values of $h$ beyond the admissible interval. First we consider the  state $\om_{\infty}$ obtained as the limit
$$
\om_{\infty}=\lim\limits_{h\to 0}\om_{h}
$$
Obviously
\begin{equation}
\om_{\infty}(W_{f})=\W{1}{f=0}{0}{f\neq 0}\label{tracestate}
\end{equation}
This state is a unique trace - state on the algebra $\Wu(\Hb,\sigma)$ i.e. $\om_{\infty}(AB)=\om_{\infty}(BA)$ for all $A,B\in \Wu(\Hb, \sigma)$. However, the state $\om_{\infty}$ is non - regular.
\par
Consider now rescaled quasi free state for the scaling parameter
$$
h>\inf\,\mr{spec}(A)
$$
 It turns out that the
quasi - free state $\om$  on the CCR algebra constructed using
symplectic form $\sigma$ can be positive functional with respect to
the multiplication given by $\sigma_{h}$ only after restriction to
some subalgebra of CCR algebra $\Wu(\Hb,\, \sigma)$. The restriction
is constructed as follows. Let $E$ be the spectral measure of the
operator $A$ which defines the state $\om$. Let
\begin{equation}
E_{h}=E((h,\, h_{\ast}]),\quad
h_{\ast}=||A||_{\mr{op}}
\end{equation}
Take $\Hb_{h}=E_{h}\Hb$ and consider the restriction of the operator
$A$ to this subspace
\begin{equation}
A^{(h)}=E_{h}AE_{h}
\end{equation}
Then
\begin{equation}
\frac{1}{h}A^{(h)}\geq \I
\end{equation}
so the quasi - free state $\om^{(h)}$  defined by this operator,  equals to the restriction of $\om$ to the sub - algebra $\Wu(\Hb_{h},\sigma)$.
Moreover, it is the quasi - free state on the algebra $\Wu (\Hb_{h},\,\sigma_{h})$. In the case of KMS state, $\inf\,\mr{spec}(A)=1$ and
$$
h_{\ast}=\frac{e^{\beta\epsilon}+1}{e^{\beta \epsilon }-1}
$$
Let $P$ be the
spectral measure of the operator $\Delta$ (see formula
 (\ref{Delta})). Since
\begin{equation}
E((h,\, h_{\ast}])=P([e^{\epsilon},\, \lambda_{\ast}))
\end{equation}
where
$$
\lambda_{\ast}=\left(\frac{h+1}{h-1}\right)^{1/\beta}
$$
The restriction $\Delta^{(h)}$ of the operator $\Delta$ to the
Hilbert space $\Hb_{h}$ is given by
$$
\Delta^{(h)}=\int\limits_{[e^{\epsilon},\lambda_{\ast})}\lambda\,dP(\lambda)
$$
Since
$$
||\Delta^{(h)}||_{\mr{op}}=\left(\frac{h+1}{h-1}\right)^{1/\beta}
$$
the operator $\Delta^{(h)}$ is bounded. Now we have
\begin{thm}
Let $h\in (1,h_{\ast})$. The restriction $\om^{(h)}$ of KMS state
$\om$ to the subalgebra $\Wu(\Hb_{h},\sigma)$ is KMS state with
respect to the time evolution defined by unitary group
$(\Delta^{(h)})^{it}$ with bounded generator.
\end{thm}
Similarly, the restriction of the operator $\Delta_{h}$ defined by
the formula (\ref{Deltah}), can be written as
\begin{equation}
\Delta_{h}^{(h)}=\int\limits_{[e^{\epsilon},\lambda_{\ast})}j_{h}(\lambda)dP(\lambda)\label{Deltahrest}
\end{equation}
Define also the unitary operators giving the time evolution
\begin{equation}
\widetilde{T}_{t}^{(h)}=(\Delta_{h}^{(h)})^{it}
\end{equation}
and
\begin{equation}
\widetilde{\alpha}_{t}^{(h)}(W_{f})=W_{\widetilde{T}_{t}^{(h)}f}
\end{equation}
 Using the similar method as in subsection \textbf{IV. B}, we can show
the following result:
\begin{thm}
Let $h\in (1,\, h_{\ast})$ and  $\om^{(h)}$ is KMS state on the
subalgebra $\Wu (\Hb_{h},\, \sigma)$. Then the rescaled state
$\om^{(h)}_{h}$ is  $(\widetilde{\alpha}_{t}^{(h)},\, \beta)$ - KMS
state on the algebra $\Wu (\Hb_{h},\,\sigma)$.
\end{thm}
This result can be formulated in the following way. Let $\cA^{(h)}=\Wu(\Hb_{h},\sigma)$. Notice that for $h^{\prime}>h>1$
$$
\cA^{(h^{\prime})}\subset \cA^{(h)}\quad\text{and}\quad \cA^{(h_{\ast})}=\{\I\}
$$
Instead of state $\om_{h}^{(h)}$ of subalgebras $\cA^{(h)}$, consider its extension $\widetilde{\om}_{h}$ to the whole algebra $\Wu(\Hb,\sigma)$, defined as
\begin{equation}
\widetilde{\om}_{h}(W_{f})=\W{\om_{h}^{(h)}(W_{f})}{f\in \Hb_{h}}{0}{\text{otherwise}}\label{nonregular}
\end{equation}
The functional (\ref{nonregular}) is $\sigma$ - positive, so it defines a state on $\Wu(\Hb, \sigma)$, but this state is non - regular. In this way we obtain the family of non - regular states of Weyl algebra $\Wu(\Hb,\sigma)$, which are regular on subalgebras $\cA^{(h)}$. Notice also that
\begin{equation}
\lim\limits_{h\to h_{\ast}}\widetilde{\om}_{h}=\om_{\infty}
\end{equation}
where $\om_{\infty}$ is a trace - state defined by (\ref{tracestate}).
\section{Conclusions}
In the present work we have studied outcomes of the variability of the Planck's constant in the quantum theory of infinite
systems. The strict formulation of the problem within the algebraic quantum theory  allowed to obtain new precise bounds 
on the changes of the $\hbar$ and interesting conclusions concerning the rescaled states. At first we have given  the 
description of isomorphisms of $C^*$-algebras  $\Wu(\Hb,\sigma_{h})$ for differing Planck's constants  and characterization 
of states in
terms of the rescaling, what provided the tool to study the influence of rescaling on states. Then,  the quasi-free states
were discussed. It turns out that such a state $\omega$ defined by the operator $A\geq \I$ remains quasi-free state on  all
algebras $\Wu(\Hb,\sigma_{h})$, for $h\in (0,\,h_{\mr{max}})$, with $h_{\mr{max}}=\inf\, \mr{spec}(A)$. The
Fock-states after the rescaling $\omega_{0,h}$ do not become the new Fock-states, and even more, are not
quasi-equivalent to the original Fock-state $\omega_{0}$. However, the rescaled Fock-states $\omega_{0,h}$ form the set
of extreme points of the convex set of universally invariant states.  The  interesting situation we have for the rescaled
KMS-states, namely,  they remain of the KMS-class for the same value of temperature ($\beta$-parameter), but for
nontrivially modified unitary evolution $\alpha^{h}_{t}$.  In the Sec. V. we analyse what happens when states are
reparametrized with values of parameter from the outside the allowed set $(0, h_{max})$, $h_{max}=\inf\,\mr{spec} (A)$. 
Such quasi-free states $\omega^{(h)}$ are non-regular, however can be interpreted as still quasi-free states on the appropriate
restriction of the CCR-algebra. Similar conclusion for such 'unproper' reparametrization is valid for the reparametrized
KMS-states, provided the relevant evolution is appropriately adapted.

In conclusion, the algebraic description of infinite quantum systems yields some restrictions for the variability of the Planck's
constant and allows the comparison of states for different physical realities defined by its actual value.

% % % % % % % % % % % % % % % % % % % % % % % % % % %

\end{document}